\documentclass[letterpaper,twocolumn,english,prb]{revtex4}
\usepackage[T1]{fontenc}
\usepackage[latin1]{inputenc}
\makeatletter

\usepackage{babel}
\makeatother
\begin{document}

\title{Compton scattering beyond the impulse approximation.}

\author{I. G. Kaplan$^{1,2*}$, B. Barbiellini$^{2}$ and A. Bansil$^{2}$}

\begin{abstract}
We treat the non-relativistic Compton scattering process in which
an incoming photon scatters from an N-electron many-body state to
yield an outgoing photon and a recoil electron, without invoking the
commonly used frameworks of either the impulse approximation (IA)
or the independent particle model (IPM). An expression for the associated
triple differential scattering cross section is obtained in terms
of Dyson orbitals, which give the overlap amplitudes between the $N$-electron
initial state and the $(N-1)$ electron singly ionized quantum states
of the target. We show how in the high energy transfer regime, one
can recover from our general formalism the standard IA based formula
for the cross section which involves the ground state electron momentum
density (EMD) of the initial state. Our formalism will permit the
analysis and interpretation of electronic transitions in correlated
electron systems via inelastic x-ray scattering (IXS) spectroscopy
beyond the constraints of the IA and the IPM. 
\end{abstract}

\address{$^{1}$Instituto de Investigaciones en Materiales,UNAM, Apdo. Postal
70-360, 04510 Mexico, D. F., Mexico. }

\address{$^{2}$Department of Physics, Northeastern University, Boston, MA
02115.}

\maketitle

\section{Introduction}

Compton scattering is unique among spectroscopic techniques in that
it allows direct experimental access to the ground state electron
momentum density (EMD) $\rho({\mathbf{p}})$ of the target many body
system \cite{5,6}. Recent high resolution Compton scattering studies
using synchrotron light sources have revealed interesting electron
correlation effects in a number of materials \cite{9,10,11,12,13,al1,be3,14,15,16}.
The experimental work has been concentrated largely on the measurement
of the double differential scattering cross section (for detecting
energy transfer and solid angle of the outgoing photon), which yields
the so-called Compton profile (CP) related to the two-dimensional
(2D) integral of the EMD \begin{equation}
J(p_{z})=\int\!\!\!\int\rho({\mathbf{p}})dp_{x}dp_{y},\label{eq1}\end{equation}
 or equivalently a one-dimensional projection of the EMD along the
direction of the scattering vector $p_{z}$ of the incident photon.

Form (\ref{eq1}) which is used in much of the existing analysis of
CP's is obtained within the framework of the impulse approximation
(IA) \cite{platzman1,7}. The fundamental scattering process considered
in the IA is the scattering of a photon from a collection of free
electrons. The IA is expected to be valid when the energy transferred
in the scattering process is much larger than the binding energy of
the electronic states involved. By its very nature, Eq. (\ref{eq1})
lacks a systematic way of taking account of deviations from the IA.
With this motivation, our purpose in this article is to consider the
general scattering event in which the incoming photon is scattered
from a bound many-electron system. We evaluate the resulting partial
triple differential scattering cross section rigorously in terms of
the so-called Dyson orbitals, which involve overlap of the $N$-body
initial state wave function with the $(N-1)$ body wave function of
the singly ionized final state with an ejected electron. The physically
relevant triple differential scattering cross section is then obtained
by summing over final (ionic) states and the steps necessary to recover
the IA are clarified. By going beyond the IA, our study provides a
systematic scheme for understanding electronic structure and correlation
effects via inelastic x-ray scattering (IXS), away from the deeply
inelastic regime.

In this connection, it is important to recognize that the standard
Compton scattering experiment does not involve the measurement of
the kinematics of the outgoing (recoil) electron. This is the reason
for the appearance of the 2D integral and the concomitant loss of
information about $\rho(\mathbf{p})$ in Eq. (\ref{eq1}). As was
pointed out first by Kaplan and Yudin \cite{17} on the basis of their
theoretical studies \cite{18} of Compton scattering on bound electrons
of light atoms and molecules, the full three-dimensional (3D) EMD
can be determined, if the characteristics of the scattered photon
and the ejected electron are measured in coincidence The authors \cite{17,18}
have also shown that, if the ejected electrons are selected by energy,
the EMD associated with individual electronic states can in principle
be obtained.

Although coincidence experiments were undertaken quite early \cite{19,20},
results for 3D EMD were first reported by Bell and collaborators \cite{21,22,23,24,25,26};
see also related work of Itoh and collaborators \cite{27,28}. Since
the cross section for an incident photon to scatter into an outgoing
electron and a photon is measured, such a measurement is often referred
to as a $(\gamma,e\gamma)$ experiment. From a formal viewpoint, the
$(\gamma,e\gamma)$ experiment provides a measurement of the triple
differential scattering cross section, for which we present in this
article a rigorous many-body expression .

For interpreting experimental CP's using the IA based formula (\ref{eq1}),
actual computations in the literature largely employ the independent
particle model (IPM). The many-electron wave function underlying the
IPM is built from Slater determinants of single-electron orbitals
obtained usually via unrestricted Hartree-Fock (UHF) approach or various
versions of the density functional theory (DFT). With a growing interest
in applying IXS for investigating electronic transitions in highly
correlated systems using synchrotron light sources, we should keep
in mind that one will need to take account of deviations not only
from the IA but also from the IPM. Even in the relatively simple case
of Li, substantial deviations in the EMD predicted by the local density
approximation have been implicated in explaining the observed discrepancies
between the computed and measured CP's \cite{38}.

Concerning other relevant literature related to the issue of going
beyond the IA, several studies have considered the accuracy of the
IA in describing core Compton profiles \cite{atom1,atom2,atom3,atom4,atom5,atom6}.
A general method for introducing final-state-interaction effects has
been discussed by Sears \cite{sears} in the context of deep-inelastic
neutron scattering. This work also discusses the Björken-scaling and
y-scaling properties of the IA which have been particularly useful
in particle physics. Recently, high resolution valence Compton profiles
(CP's) of Li at a relatively low photon energy of 8-9 KeV were considered
in Refs. \cite{sternemann,soininen}. The observed asymmetries in
shape and smearing of the Fermi surface features in the CP's were
attributed to the breakdown of the IA. It is further shown that these
discrepancies in Li can be understood in terms of a finite width of
the final state spectral function%
\footnote{The approach of Refs.\cite{sternemann,soininen} considers the dynamical
structure factor in terms of a product of electron and hole single
particle spectral functions. The relationship with our approach can
be understood by noting that the Dyson orbital considered in this
article is related to the hole wavefunction while the ejected electron
goes in a high energy continuous state.%
}. To our knowledge, all previous work concerning the breakdown of
the IA has been dedicated to understanding the double differential
scattering cross section. The present study focuses on the elementary
$(\gamma,e\gamma)$ scattering process and provides a clearer picture
of the many-body effects and their connection with the IA and the
IPM.

An outline of this article is as follows. These introductory remarks
are followed in Section \ref{ssa} with a rigorous treatment of the
partial triple differential scattering cross section in terms of Dyson
orbitals. Section \ref{sub:ssb} addresses the question of summing
over final states to obtain the total triple differential cross section
and how it reduces to the IA result proportional to the EMD. Section
III makes a few concluding remarks.

\section{General expressions for triple differential scattering cross section}

\subsection{Partial triple differential cross section}

\label{ssa} We consider the non-relativistic elementary scattering
process in which the incoming photon of energy $\omega_{1}$ (here
and throughout this article, natural units, $\hbar=c=1$, are assumed
implicitly) and momentum $\mathbf{k}_{1}$ scatters from the N-electron
many body ground state of the solid with energy $E_{0}(N)$. The final
state consists of: an outgoing photon with energy $\omega_{2}$ and
momentum $\mathbf{k}_{2}$; the $(N-1)$ electron ionized state of
the solid characterized by quantum number $n$ and energy $E_{n}(N-1)$;
a recoil electron carrying kinetic energy $E_{e}^{(n)}$ and momentum
$\mathbf{p}_{n}$. We assume that the momentum transferred to the
ionic system is $\mathbf{q}$ and the associated kinetic energy is
neglected given the large mass of the target. The scattering process
is illustrated schematically in Fig. 1(a). The total momentum ${\mathbf{k}}$
transferred through the scattering of the photon is \begin{equation}
{\mathbf{k}}=\mathbf{k}_{1}-\mathbf{k}_{2}=\mathbf{q}+\mathbf{p}_{n}~,\label{eq3}\end{equation}
 where the second equality gives the momentum conservation condition,
which is shown also in Fig. 1(b) %
\footnote{An additional electron can also be knocked off from the ground state
producing a doubly ionized state, see J.H. McGuire, J. Wang, and J.
Burgdrfer, Phys. Rev. Lett.\textbf{77}, 1723 (1996). Multiple ionization
processes are not consider here.%
}.

Care is needed in formulating the energy conservation condition. In
the standard treatment of the Compton scattering process, one assumes
independent electrons with various one-particle energies. However,
the preceding discussion makes it clear that the general interacting
system is more naturally characterized via the quantum number $n$
of the ionized target. Therefore the relvant binding energy $E_{b}^{(n)}$
is \begin{equation}
E_{b}^{(n)}=E_{n}(N-1)-E_{0}(N)\label{eq2}\end{equation}
 in terms of the ground state energy of the N-particle system and
that of the $(N-1)$ particle ionized target. In the one-particle
approximation, $E_{b}^{(n)}$ will correspond to the energy of the
orbital from which the outgoing electron is ejected. The energy conservation
then yields \begin{equation}
\omega_{1}-\omega_{2}=E_{b}^{(n)}+E_{e}^{(n)}~.\label{eq2ter}\end{equation}

In addition to energy and momentum, the total spin is also conserved
in the scattering process. If the target is initially in an $S=0$
state, then the final state will also be a singlet, while for a magnetized
sample in a triplet state, the final state can be a doublet or a quartet.

As shown in Ref. \cite{7}, in the high energy transfer region, where
$\omega_{1}-\omega_{2}>>E_{b}^{(n)}$, the interaction between the
electromagnetic field and the target can be approximated by %
\footnote{The linear interaction term in $\mathbf{A}$ appears only in the second
order cross section formula and can be neglected in the absence of
resonances.%
}. \begin{equation}
V_{int}=\frac{e^{2}}{2mc^{2}}A^{2}~,\label{eq4}\end{equation}
 where $\mathbf{A}$ is the vector potential of the field. Using this
form of the interaction, the expression for the double differential
Compton scattering cross section in the IA was obtained in Ref. \cite{7}.
The IA corresponds in effect to modeling the scattering process as
an elastic collision between a photon and an electron of a particular
momentum with the target being represented by a distribution of such
independent electronic states.

More relevant for our purposes is the treatment of Refs. \cite{17,18}.
These authors obtain the cross section for the elementary Compton
scattering process (in which the ion is left behind in a specific
quantum state) from a many-body molecular system in the nonrelativistic
$A^{2}$ approximation of Eq. (\ref{eq4}). It is natural to refer
to such a cross section as a \textit{partial} triple differential
scattering cross section (PTDSC), since the total triple differential
scattering cross section (TTDSC) is obtained by summing the PTDSC
over final states (see Section \ref{sub:ssb} below). The expression
for the PTDSC is \cite{17,18}

\begin{equation}
\frac{d^{3}\sigma_{n}}{d\omega_{2}d\Omega_{2}d\Omega_{e}}=\frac{r_{0}^{2}}{2}(1+\cos^{2}\theta)\frac{\omega_{2}}{\omega_{1}}|M^{(n)}|^{2}\delta(\omega_{1}-\omega_{2}-E_{b}^{(n)}-E_{e}^{(n)})~,\label{eq:5}\end{equation}
 where $\delta$ denotes the Kronecker symbol (here and elsewhere
in this article) and reflects the energy conservation law in Compton
scattering, $r_{0}=e^{2}/mc^{2}$ is the classical electron radius,
$\theta$ is the scattering angle, and \begin{equation}
M^{(n)}=<\Psi_{f}^{(n)}(\mathbf{x}_{1}....\mathbf{x}_{N})~|~\sum_{\nu=1}^{N}\exp(i{\mathbf{k}}\mathbf{r}_{\nu})~|~\Psi_{i}(\mathbf{x}_{1}....\mathbf{x}_{N})>\label{eq6}\end{equation}
 is the transition matrix element calculated with $N$-electron wave
functions of initial and final states of the target. Note that expression
(\ref{eq:5}) for the PTDSC assumes an implicit summation over the
vibrational states within the framework of the Born-Oppenheimer approximation;
in any event, it will be difficult to resolve vibrational levels in
the Compton scattering regime \cite{18,41}.

The antisymmetry of the many electron wavefunction implies that the
contribution from each term in expression (\ref{eq6}) is the same.
Therefore, we may replace the summation over $\nu$ by the last term,
which corresponds to the ejection of the $N$-th electron, yielding
\begin{equation}
M^{(n)}=N<\Psi_{f}^{(n)}(\mathbf{x}_{1}....\mathbf{x}_{N})~|~\exp(i{\mathbf{k}}\mathbf{r}_{N})~|~\Psi_{i}(\mathbf{x}_{1}....\mathbf{x}_{N})>~.\label{eq7}\end{equation}

We assume now that the initial state possesses a total spin $S=0$,
as is the case in most nonmagnetic materials. The final state will
then be a singlet state (due to spin conservation in the Compton scattering
process) and the associated antisymmetric singlet wave function can
be represented as \begin{eqnarray}
\Psi_{f}^{(n)}(\mathbf{x}_{1}....\mathbf{x}_{N}) & = & \hat{\mathcal{A}}\frac{1}{\sqrt{2}}[\Psi_{n\alpha}(\mathbf{x}_{1}....\mathbf{x}_{N-1})\psi_{{\mathbf{p}}\beta}(\mathbf{x}_{N})\nonumber \\
 &  & -\Psi_{n\beta}(\mathbf{x}_{1}....\mathbf{x}_{N-1})\psi_{{\mathbf{p}}\alpha}(\mathbf{x}_{N})]~.\label{eq8}\end{eqnarray}
 Here $\psi_{{\mathbf{p}}\sigma}(\mathbf{x}_{N})$ is the wave function
of the ejected electron with momentum $\mathbf{p}_{n}$ and spin projection
$\sigma$ that can accept only two values: $\sigma=1/2$ denoted by
$\alpha$ and $\sigma=-1/2$ by $\beta$. $\Psi_{n\alpha}$ and $\Psi_{n\beta}$
are the two doublet components of the $(N-1)$-electron ionic wave
function in the $n$-th quantum state (in the one-electron picture,
this describes an ion with a hole in the $n$-th shell). The ionic
states are the eigenstates of the $(N-1)$-electron Hamiltonian. $\hat{\mathcal{A}}$
is an antisymmetrization operator given by \begin{equation}
\hat{\mathcal{A}}=\frac{1}{\sqrt{N}}(1-\sum_{\nu=1}^{N-1}P_{\nu N})~,\label{eq9}\end{equation}
 where the permutation $P_{\nu N}$ transposes the ejected $N$th
electron with the $\nu$th electron in the ion.

Substituting the final state wave function (\ref{eq8}) into Eq. (\ref{eq7})
and invoking the condition of strong orthogonality

\begin{equation}
\begin{array}{c}
\left\langle [\Psi_{n\alpha}(\mathbf{x}_{1}....\mathbf{x}_{N-1})\psi_{{\mathbf{p}}\beta}(\mathbf{x}_{N})-\right.\\
\Psi_{n\beta}(\mathbf{x}_{1}....\mathbf{x}_{N-1})\psi_{{\mathbf{p}}\alpha}(\mathbf{x}_{N})]\left|\Psi_{i}(\mathbf{x}_{1}....\mathbf{x}_{N})\right\rangle =0\end{array}\label{eq:10}\end{equation}
 for all $i$, we obtain

\begin{equation}
\begin{array}{ccc}
M^{(n)}= & \sqrt{\frac{N}{2}}<[\Psi_{n\alpha}(\mathbf{x}_{1}....\mathbf{x}_{N-1})\psi_{{\mathbf{p}}\beta}(\mathbf{x}_{N})\\
 & -\Psi_{n\beta}(\mathbf{x}_{1}....\mathbf{x}_{N-1})\psi_{{\mathbf{p}}\alpha}(\mathbf{x}_{N})]~|~\exp(i{\mathbf{k}}\mathbf{r}_{N})~\\
 & |~\Psi_{i}(\mathbf{x}_{1}....\mathbf{x}_{N})>~.\end{array}\label{eq:11}\end{equation}
 This can be represented in terms of the Dyson spin-orbitals \cite{42,43,43bis,44,45},
defined by \begin{equation}
g_{n}(\mathbf{x}_{N})=\sqrt{N}\int~\Psi_{n}(\mathbf{x}_{1}....\mathbf{x}_{N-1})^{\ast}\Psi_{0}(\mathbf{x}_{1}....\mathbf{x}_{N})~d\mathbf{x}_{1}....d\mathbf{x}_{N-1}~,\label{eq12}\end{equation}
 where the integration over $d\mathbf{x}_{i}$ includes a summation
over the spin coordinates. The Dyson spin-orbitals $g_{n}(\mathbf{x}_{N})$
may thus be thought of as generalized overlap amplitudes between the
ground state and the ionized states of the many body system. They
naturally appear in the spectral resolution of the one-particle Green
function \cite{46,47}, and have been exploited successfully in some
studies of ionization of atomic and molecular systems by electromagnetic
radiation or fast electrons \cite{del1,del2,del3}. Note that in general
Dyson orbitals do not form an orthonormal set. Some authors \cite{45}
define Dyson orbitals without the pre-factor of $\sqrt{N}$. The Dyson
spin-orbital with the spin projection $\sigma$ may be written in
terms of the spin function $\sigma(\zeta)$ as \begin{equation}
g_{n}(\mathbf{x}_{N})=g_{n}(\mathbf{r}_{N},\sigma({\zeta_{N}}))=g_{n}(\mathbf{r}_{N})\sigma(\zeta_{N})~.\label{eq13}\end{equation}
 The wave function of the ejected electron similarly is \begin{equation}
\psi_{\mathbf{p}_{n}\alpha}(\mathbf{x}_{N})=\psi_{\mathbf{p}_{n}}(\mathbf{r}_{N})\alpha(\zeta_{N})~.\label{eq14}\end{equation}

Introducing definition (\ref{eq12}) into Eq. (\ref{eq:11}) and performing
spin integration, we obtain a compact general expression for the transition
matrix element \begin{equation}
M^{(n)}=\sqrt{2}\int g_{n}({\mathbf{r}})\exp(i{\mathbf{k}}{\mathbf{r}})\phi_{\mathbf{p}_{n}}^{*}({\mathbf{r}})~d{\mathbf{r}}~.\label{eq15}\end{equation}

In the region of large energy transfer, Eq. (\ref{eq15}) provides
an exact expression for the matrix element in terms of the Dyson orbital
$g_{n}(\mathbf{x}_{N})$ and the wave function of the ejected electron
in the potential field of the ion. Electron correlation effects enter
through $g_{n}(\mathbf{x}_{N})$ and can be included in any particular
scheme to the extent to which these are incorporated in the computation
of this quantity. In general, Dyson orbitals can be expanded into
linear combinations of Hartree-Fock or other one-particle wave functions.
In the so-called diagonal approximation, the Dyson orbital is equal
to the square root of the pole strength times the HF orbital \cite{43bis,45},
and can be calculated using special code \cite{48} implemented into
the Gaussian-98 program suite \cite{49}; see also Ref. \cite{50}.

Under the condition $\omega_{1}-\omega_{2}>>E_{b}^{(n)}$, the wave
function of ejected electron $\phi_{\mathbf{p}_{n}}({\mathbf{r}})$
may be approximated as a plane wave \begin{equation}
\phi_{\mathbf{p}_{n}}({\mathbf{r}})=\frac{1}{(2\pi)^{3/2}}\exp(i\mathbf{p}_{n}{\mathbf{r}})~,\label{eq16}\end{equation}
 allowing the transition matrix element to be expressed via the Dyson
orbital $g_{n}({\mathbf{q}})$ in momentum space \begin{equation}
M^{(n)}=\frac{1}{2\pi^{3/2}}\int g_{n}({\mathbf{r}})\exp(i{\mathbf{q}}{\mathbf{r}})~d{\mathbf{r}}=\sqrt{2}~g_{n}({\mathbf{q}}),\label{eq17}\end{equation}
 Here, ${\mathbf{q}}={\mathbf{k}}-\mathbf{p}_{n}$ is the momentum
transferred to the ion. Since the ejected electron is considered as
being free (with energy $p_{n}^{2}/2m$), the absolute value of the
vector $\mathbf{p}_{n}$ is completely determined by the energy conservation
law and is equal to \begin{equation}
p_{n}=\sqrt{2m(\omega_{1}-\omega_{2}-E_{b}^{(n)})}~.\label{eq18}\end{equation}
 The direction of the vector $\mathbf{p}_{n}$ is undetermined so
that only the maximum and minimum values of the vector ${\mathbf{q}}$
are constrained as follows \begin{equation}
|k-p_{n}|\leq q\leq k+p_{n}~.\label{eq19}\end{equation}
 In this sense, vector ${\mathbf{q}}$ involves an implicit dependence
on the index $n$.

Using Eqs. (\ref{eq16}) and (\ref{eq:5}), we obtain the final expression
for the PTDSC where the ion is created in a definite electronic state
$n$ as \begin{equation}
\frac{d^{3}\sigma_{n}}{d\omega_{2}d\Omega_{2}d\Omega_{e}}=r_{0}^{2}(1+\cos^{2}\theta)\frac{\omega_{2}}{\omega_{1}}|g_{n}({\mathbf{q}})|^{2}\delta(\omega_{1}-\omega_{2}-E_{b}^{(n)}-\frac{p_{n}^{2}}{2m})~.\label{eq20}\end{equation}
 Note that here the independent particle approximation (IPM) is not
invoked. In the IPM, the Fourier component of the Dyson orbital in
Eq. (\ref{eq20}) reduces to the Fourier component of the Hartree-Fock
or the Kohn-Sham orbital from which the electron is removed. Moreover,
aside from the use of the plane wave form (\ref{eq16}) for the ejected
electron wave function, expression (\ref{eq20}) does not invoke the
impulse approximation.

The determination of the PTDSC and $g_{n}(\mathbf{q})$ in Eq. (\ref{eq20})
requires measurements of the angular and energy characteristics of
both the scattered photon and the ejected electron taken in coincidence.
In order to understand the relevant experimental geometries, it is
helpful to refer to the momentum conservation condition depicted in
Fig. 1b above. Kaplan and Yudin \cite{17} suggested a scheme in which
the characteristics of the outgoing photon beam (i.e. the angle $\theta$
and energy $\omega_{2}$ in Fig. 1(a)) are fixed, but the angle $\theta_{e}$
of the ejected electron is varied to access different $\mathbf{q}$-values.
The fixed value of $\omega_{2}$ should be selected near the peak
of the Compton line \cite{17,18} (i.e. close to the value given by
the Compton formula for the scattering from a free electron at rest).
By measuring the energy of the ejected electron then, one can, in
principle, select the specific quantum state $n$ involved in the
scattering process through the energy conservation condition (\ref{eq2ter}).
Another approach, followed more recently by Itoh and collaborators,
is to fix the position of the electron as well as the photon detector
(i.e. the angles $\theta$ and $\theta_{e}$ in Fig. 1(a)), but energy
analyze both the scattered photon and the recoil electron in coincidence
\cite{27,28}.

\subsection{Summation over final states\label{sub:ssb}}

The TTDSC is obtained from Eq. (\ref{eq20}) by summing over the available
final states $n$ as %
\footnote{The occupied Dyson orbitals $g_{n}$ are defined via the overlap of
the $N$-electron ground state wavefunction and the $(N-1)$-electron
ionized states. Similary one can define unoccupied Dyson orbitals
$f_{n}$ by considering states of the system with an additional added
electron. %
}

\begin{eqnarray}
\frac{d^{3}\sigma}{d\omega_{2}d\Omega_{2}d\Omega_{e}}=r_{0}^{2}(1+\cos^{2}\theta)\frac{\omega_{2}}{\omega_{1}}\nonumber \\
\times\sum_{n}|g_{n}({\mathbf{q}})|^{2}\delta(\omega_{1}-\omega_{2}-E_{b}^{(n)}-\frac{p_{n}^{2}}{2m})~.\label{eq21}\end{eqnarray}
 In the high-energy transfer region, $\omega_{1}-\omega_{2}>>E_{b}^{(n)}$,
the binding energy in the Kronecker $\delta$-function on the right
hand side may be neglected, so that the absolute value of momentum
$\mathbf{p}_{n}$ in Eq. (\ref{eq18}) becomes independent of $n$
and the summation over $n$ simplifies to yield \begin{eqnarray}
\sum_{n}|g_{n}({\mathbf{q}})|^{2}\delta(\omega_{1}-\omega_{2}-E_{b}^{(n)}-\frac{p_{n}^{2}}{2m}) & =\nonumber \\
\delta(\omega_{1}-\omega_{2}-\frac{p_{e}^{2}}{2m})\sum_{n}|g_{n}({\mathbf{q}})|^{2}~,\label{eq22}\end{eqnarray}
 where $\mathbf{p}_{n}$ is replaced by $\mathbf{p}_{e}$ to emphasize
that the momentum of the outgoing electron is independent of state
$n$.

The sum of $|g_{n}({\mathbf{q}})|^{2}$ over all occupied states $n$
can be expressed via the one-particle reduced density matrix \cite{51,52,53}
for an $N$-electron system defined as 

\begin{equation}
\begin{array}{c}
\Gamma_{1}({\mathbf{r}};\mathbf{r}^{\prime})=N\int~\Psi(\mathbf{x}_{1}....\mathbf{x}_{N-1},{\mathbf{r}},\zeta)^{*}\Psi(\mathbf{x}_{1}....\mathbf{x}_{N-1},\mathbf{r}^{\prime},\zeta)\\
\times d\mathbf{x}_{1}....d\mathbf{x}_{N-1}d\zeta.\end{array}\label{eq23}\end{equation}
 In the momentum space \begin{equation}
\Gamma_{1}({\mathbf{q}};{\mathbf{q}})=\frac{1}{(2\pi)^{3}}\int~\Gamma_{1}({\mathbf{r}},\mathbf{r}^{\prime})\exp(i{\mathbf{q}}({\mathbf{r}}-\mathbf{r}^{\prime}))~d{\mathbf{r}}~d\mathbf{r}^{\prime}~.\label{eq24}\end{equation}

We now recall the following decomposition of the one-particle reduced
density matrix \cite{47}\begin{equation}
\Gamma_{1}({\mathbf{r}};\mathbf{r}^{\prime})=\sum_{n}g_{n}({\mathbf{r}})g_{n}(\mathbf{r}^{\prime})^{*}~.\label{eq25}\end{equation}
 Substituting this decomposition into Eq. (\ref{eq24}), one obtains
\begin{equation}
\sum_{n}|g_{n}({\mathbf{q}})|^{2}=(2\pi)^{3}\Gamma_{1}({\mathbf{q}};{\mathbf{q}})\equiv(2\pi)^{3}\rho({\mathbf{q}})~.\label{eq26}\end{equation}
 Thus, in the high-energy transfer region, the TTDSC is directly related
to the 3D EMD as follows \begin{equation}
\frac{d^{3}\sigma}{d\omega_{2}d\Omega_{2}d\Omega_{e}}=(2\pi)^{3}r_{0}^{2}(1+\cos^{2}\theta)\frac{\omega_{2}}{\omega_{1}}\rho({\mathbf{q}})\delta(\omega_{1}-\omega_{2}-\frac{p_{e}^{2}}{2m})~.\label{eq27}\end{equation}

It is this TTDSC that is measured in the $(\gamma,e\gamma)$ experiments
by Bell and collaborators \cite{24,25,26}, issues of experimental
resolution notwithstanding. Note however that there is an interesting
difference in the way the momentum density factor $\rho$ occurs in
Eq. (\ref{eq27}) compared to the analytical expressions employed
by Refs. \cite{24,25,26}. In our case, the EMD ($\rho(\mathbf{q})$)
is sampled at the momentum \begin{equation}
{\mathbf{q}}={\mathbf{k}}-\mathbf{p}_{e}~,\label{eq28}\end{equation}
 which is the momentum transferred to the ion, whereas in the cross
section of Refs. \cite{24,25,26,27,28}, the EMD involved is $\rho({\mathbf{p}})$,
where ${\mathbf{p}}$ is the initial momentum of the electron before
scattering. The reason is that the study of Refs. \cite{24,25,26}
is based on the formulae of Ribberfors \cite{39} for double differential
cross sections in the IA. As already noted, in the IA the scattering
is the same as for free electrons, but weighted with the probability
with which the plane-wave state of momentum ${\mathbf{p}}$ occurs
in the ground state. For a system of bound particles, this picture
does not constitute a useful starting point, and our more general
treatment indicates that the quantity that occurs naturally is the
momentum ${\mathbf{q}}$ transferred to the ion. Nevertheless, in
the IA, the two pictures are equivalent because in this regime, from
momentum conservation, one obtains: \begin{equation}
{\mathbf{p}}+\mathbf{k}_{1}=\mathbf{p}_{e}+\mathbf{k}_{2}\label{eq29}\end{equation}
 or equivalently \begin{equation}
{\mathbf{p}}=\mathbf{p}_{e}-{\mathbf{k}}~.\label{eq29bis}\end{equation}
 Comparing Eq. (\ref{eq28}) and (\ref{eq29bis}), we see that in
the high energy transfer limit, $\mathbf{q}$ and $\mathbf{p}$ differ
only by direction (they are opposite), as the outgoing electron loses
all memory of the bound state it came from. The range of $q$ in formula
(\ref{eq19}) becomes \begin{equation}
0\leq q\leq2k.\label{rangep}\end{equation}
 Thus, the maximum momentum transferred to the ion is given by $2k$.
In the IA, $2k$ may be interpreted as the highest momentum of an
electron in the initial system that can be ejected.

\section{Summary and Conclusions}

We start by considering the elementary process involved in Compton
scattering, namely, the scattering of an incoming photon from the
ground state of an $N$-electron target to yield a final state containing
a singly ionized target with $(N-1)$ electrons in a specific quantum
state together with an outgoing photon and an ejected electron. The
associated PTDSC (partial triple differential scattering cross section)
is obtained rigorously without resorting to the approximations inherent
in either the IA or the IPM. It is shown that the PTDSC can be expressed
in terms of the Dyson orbitals, which give the overlap between the
wave function of the ground state of the $N$-electron initial system
with the $(N-1)$ electron ionized final state wave function. The
TTDSC (total triple differential scattering cross section) is then
obtained by summing over final states, which is equivalent to summing
over the occupied Dyson orbitals. Interestingly, in our general treatment,
the momentum that plays a fundamental role in the formula for the
cross section is the momentum $\mathbf{q}$ transferred to the ion
in the scattering process and not the momentum $\mathbf{p}$ associated
with the electronic system as is the case in the IA based treatment.
We show how in the limiting case of the high energy transfer regime,
our formalism reduces to the standard IA description. Although our
treatment is non-relativistic, extension to the relativistic case
is straightforward by using relevant results in the literature \cite{39,rel}.Our
formalism will permit the analysis and interpretation of electronic
transitions in correlated electron systems via IXS beyond the constraints
of the IA and the IPM.

\begin{acknowledgments}
\noindent We are grateful to N. Dobrodey, J.V. Ortiz, and V. Zakrzewski
for useful discussions concerning the Dyson orbitals. This research
was supported by the project 32227-E CONACyT (Mexico) and by the US
Department of Energy under contract DE-AC03-76SF00098. 
\end{acknowledgments}

\newpage \textbf{Figure Caption}

\noindent \textbf{Figure 1:} (a) Schematic diagram of the elementary
scattering event involved in the Compton scattering process. The incoming
photon scatters from the target to produce an outgoing photon and
an electron and leaves the singly ionized target (not shown) in a
definite quantum state denoted by index $n$. The notation for kinematic
variables is obvious. (b) Momentum conservation in the process of
Fig. 1(a), where $\mathbf{q}$ is the momentum transferred to the
ionized target. 
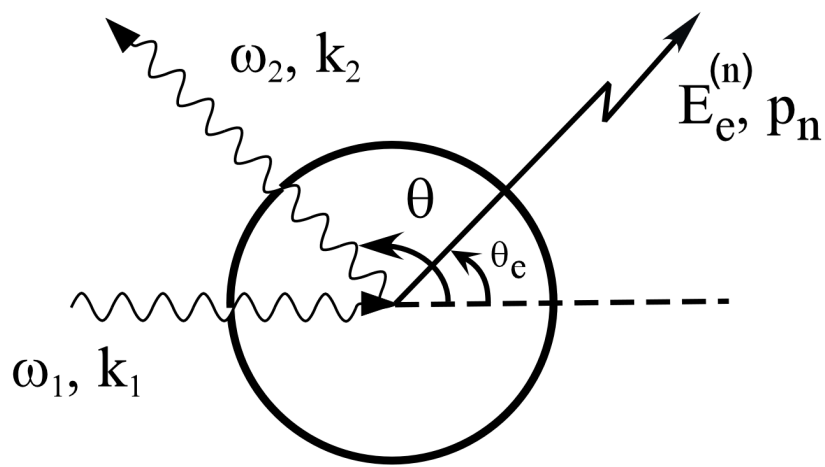
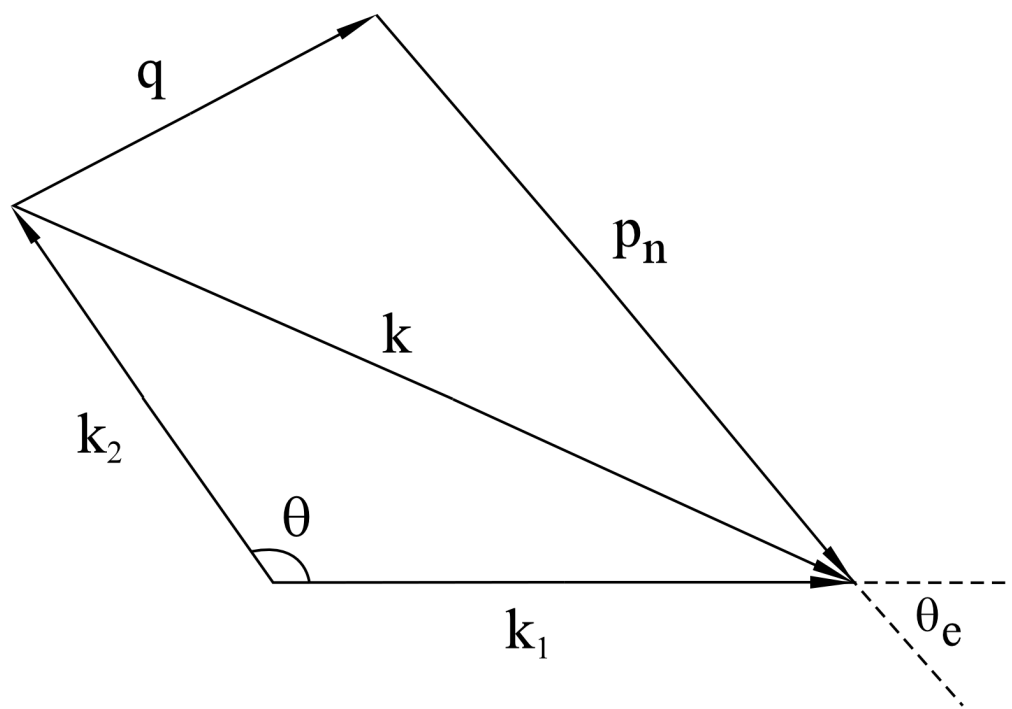
\end{document}